\documentclass{aa}
\usepackage{graphics}

\def\bmvg{$(B-V)_{o,g}$}
\def\d14{$\Delta V_{1.4}$}
\def\d12{$\Delta V_{1.2}$}
\def\d11{$\Delta V_{1.1}$}

\def\gsim{\;\lower.6ex\hbox{$\sim$}\kern-7.75pt\raise.65ex\hbox{$>$}\;}
\def\lsim{\;\lower.6ex\hbox{$\sim$}\kern-7.75pt\raise.65ex\hbox{$<$}\;}
\begin{document}

\thesaurus{05 (08.01.1; 10.07.2)} 
\title{On the calibration of Red Giant Branch metallicity indicators in
Globular Clusters: new relations based on an improved abundance scale}

\author{Eugenio Carretta, 
 \and  
 Angela Bragaglia}

\institute{Osservatorio Astronomico di Bologna, Via Zamboni 33, 
I-40126 Bologna, ITALY \\
{\emph e-mail: carretta@astbo3.bo.astro.it, angela@astbo3.bo.astro.it}}

\offprints{E. Carretta}

\date{Received  / Accepted }

\titlerunning{Photometric indices}
\authorrunning{Carretta \& Bragaglia}

\maketitle

\begin{abstract}
We present an improved calibration of photometric metallicity indicators,
derived from the new metallicity scale for Globular Clusters presented by
Carretta \& Gratton (1997) and based on direct high resolution spectroscopy
of 160 stars in 24 globular clusters.

We have carefully recalibrated the traditional abundance indices based upon
the red giant branch (RGB) morphology, both in the $V,B-V$ and $V,V-I$
planes, namely the dereddened colour at the luminosity level of the
horizontal branch (HB), and the magnitude difference between the HB and 
the RBG at a given dereddened colour.

Finally, we give new accurate relations to employ in the Simultaneous
Metallicity Reddening method by Sarajedini (1994), also tied to the
Carretta \& Gratton (1997) abundance scale.

\end{abstract}


\section{Introduction}

Since globular clusters (GCs) are among the oldest objects in galaxies, they
are widely recognized as  very useful tracers of the chemical and dynamical
evolution of their parent hosts.

The accurate knowledge of their global metal content, measured  by the [Fe/H]
ratio, is critical for many astrophysical problems. In particular, being very
massive and luminous systems of coeval stars that show, to a first
approximation, a similar (initial) chemical composition, globular clusters
represent the cornerstones in establishing the existence of an age-metallicity
relation and/or a metallicity-galactocentric distance gradient, up to the most
distant regions of the galactic halo. This in turn provides strong constraints
on models of galactic formation. Moreover, variations in the [Fe/H] content
among globular clusters can be interpreted as a fossil record of the global
processes of chemical enrichment occurred through the history of the Galaxy.
Finally, precise metallicities are one of the basic ingredients in deriving
accurate ages using parallaxes measured by the Hipparcos astrometry satellite
(see Gratton et al. 1997; Reid 1997).

Even if the best way to get a quantitatively accurate estimate of the metal
abundance of any star is detailed abundance analysis of high resolution
spectra, there are unfortunately some shortcomings that limit the application
of this technique to the study of GCs. Due to their large distances, reliable
high resolution, high signal-to-noise spectra can be obtained with the present
day instrumentation only for the brightest giants. Only an handful of stars
near the main sequence turn-off (hence reflecting the initial chemical
composition, undisturbed by mixing in later evolutionary phases) have been
observed yet. Moreover, high-resolution spectroscopy is a very time consuming
observing technique.

Therefore, in the past years, a number of indirect metallicity indicators have
been devised to overcome these problems. Almost all of them are based on
integrated parameters that bypass the distance limit also for very far
clusters, but they require a very accurate calibration in order to provide the
true content in [Fe/H]. A direct calibration, able to tie the observed
photometric indices to the actual number of iron atoms as measured from high
resolution spectral line profiles is henceforth strongly needed. In Section 2
we discuss the philosophy of our approach; Section 3 and 4 are devoted to the
presentation and discussion of new calibrations for several metallicity
indicators; a short summary is presented in Section 5.

\begin{table*}
\caption{GCs used as calibrators. Data are taken from GC97, ZW, SL (tab. 5), 
Sarajedini (1994, tab.1), and Sarajedini \& Milone (1995, for NGC5053 and 
NGC4590).
Y or N indicate in columns 3 and 9 whether the GC is a primary calibrator
for GC97 and SL respectively.}
\begin{tabular}{lcccccccccc}
\hline
&&&&&&&&&&\\
GC &[Fe/H] &Direct &[Fe/H] &E(B-V) 
   &(B-V)$_{0,g}$  &$\Delta V_{1.2}$  &$\Delta V_{1.1}$ & calib. 
   &(V-I)$_{0,g}$  &$\Delta V_{1.2}$ \\
& CG97 &spectr.? &ZW & & &$V,B-V$ &$V,B-V$ &in SL? & &$V,V-I$ \\
&&&&&&&&&&\\
\hline
&&&&&&&&&&\\
 NGC104 (47Tuc)   &-0.70 &Y &-0.71 &0.04 &0.958 &1.275 &0.798 &Y &1.032 &1.028 \\
 NGC288    &-1.07 &Y &-1.40 &0.02 &0.852 &1.884 &1.503 &N && \\
 NGC362    &-1.15 &Y &-1.28 &0.03 &0.832 &2.050 &1.741 &N && \\
 NGC1261   &-1.09 &N &-1.31 &0.00 &0.860 &2.095 &1.735 &N && \\
 NGC1851   &-1.08 &N &-1.29 &0.02 &0.873 &1.806 &1.433 &Y &0.953 &1.609  \\
 NGC1904   &-1.37 &Y &-1.69 &0.01 &0.801 &2.332 &2.001 &N && \\
 NGC4590 (M68)   &-1.99 &Y &-2.09 &0.07 &0.694 &2.810 &2.508 &Y &0.885 &2.470 \\
 NGC5053   &-2.43 &N &-2.41 &0.06 &0.647 &3.101 &2.741 &Y &0.847 &2.770 \\
 NGC6352   &-0.64 &Y &-0.60 &0.21 &0.994 &0.953 &0.591 &N && \\
 NGC6397   &-1.82 &Y &-1.91 &0.18 &0.717 &2.842 &2.483 &N &0.904 &2.330 \\
 NGC6535   &-1.53 &N &-1.75 &0.44 &0.745 &2.537 &2.134 &N && \\
 NGC6752   &-1.42 &Y &-1.54 &0.04 &0.781 &2.264 &1.873 &Y &0.949 &1.935 \\
 NGC7078 (M15)  &-2.12 &Y &-2.17 &0.10 &0.691 &2.972 &2.601 &Y &0.882 &2.538 \\
 NGC7089 (M2)  &-1.34 &N &-1.58 &&&&&                          & 0.934 &2.039\\
 Eridanus  &-1.18 &N &-1.41 &0.03 &0.838 &1.897 &1.596 &N && \\
 ESO121    &-0.83 &N &-0.93 &0.03 &0.907 &1.507 &1.101 &N && \\
 Lindsay1  &-0.94 &N &-1.10 &0.04 &0.864 &1.888 &1.522 &N && \\
 Pal14     &-1.36 &N &-1.60 &0.05 &0.803 &2.288 &1.939 &N && \\
&&&&&&&&&&\\
\hline
\end{tabular}
\end{table*}

\begin{figure*}
\hspace{3cm}\resizebox{12cm}{17cm}{\includegraphics{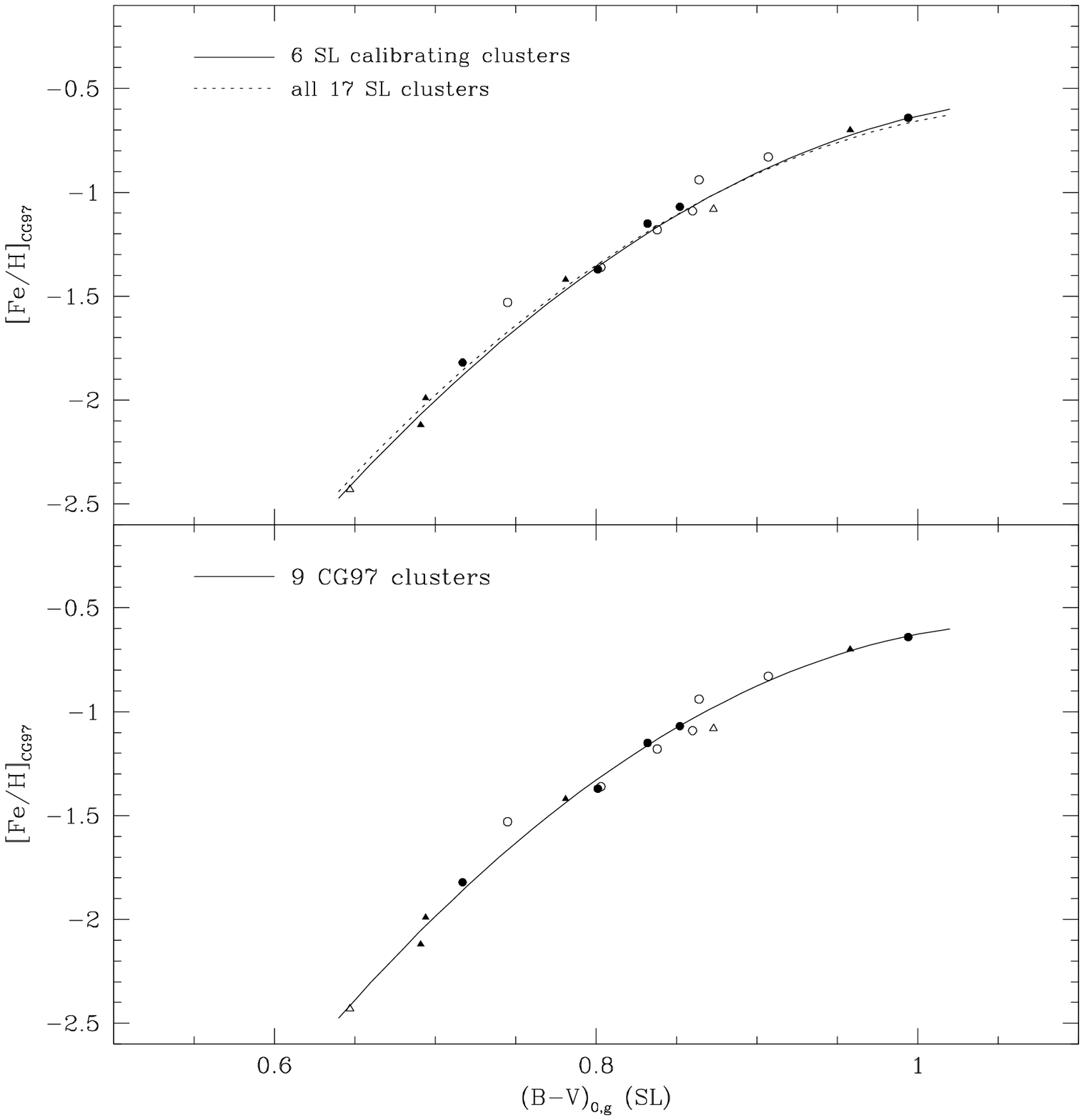}}
\caption{Calibration of the $(B-V)_{0,g}$ parameter using the CG97 metallicity
scale and the cluster sample by SL. The upper panel shows the calibrations
obtained using only the 6 SL primary calibrators (solid line) and all the 17
clusters (dotted line). In the lower panel the calibration is based only upon
the 9 clusters that have metallicities derived by CG97 from direct analysis
(CG97 reference clusters). In both panels filled symbols represent clusters
with [Fe/H]'s derived in CG97, while open symbols represent clusters with ZW
metallicities corrected to the CG97 scale. Triangles, filled or open, indicate
the 6 SL primary calibrators.}
\end{figure*}

\begin{figure*}
\hspace{3cm}\resizebox{12cm}{17cm}{\includegraphics{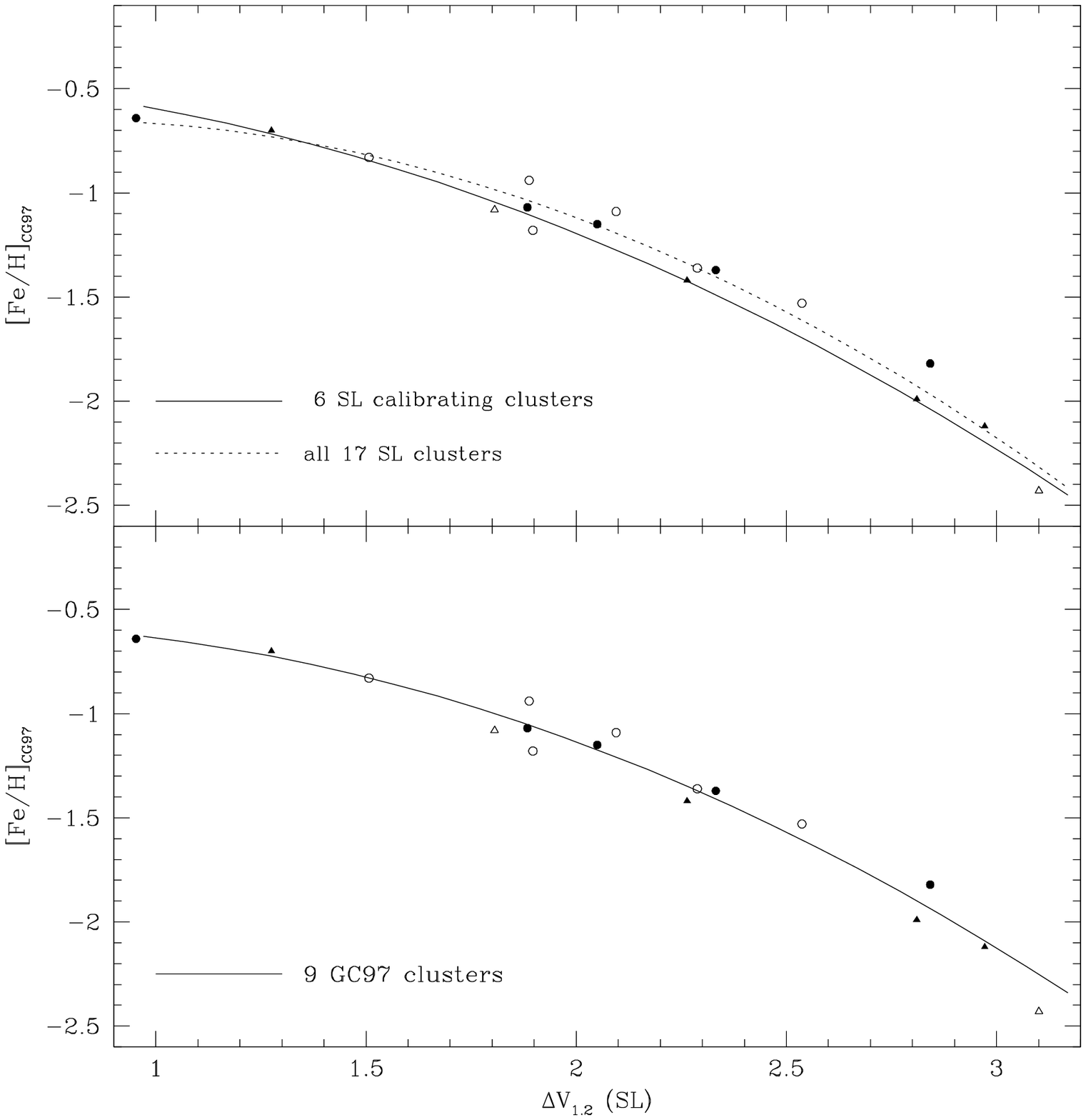}}
\caption{Calibration of the $\Delta V_{1.2}$ parameter of SL using the CG97
metallicities. The meaning of symbols is as in Figure 1.}
\end{figure*}

\section{The reference metallicity scale}

Traditionally, all indices have been tied up to now to the metallicity scale
defined by the Zinn's group (Zinn \& West 1984, Armandroff et al. 1992, Da
Costa \& Armandroff 1990; hereinafter ZW on the whole), systematically
ignoring any later result from high dispersion spectroscopy. ZW's scale is a
compilation of metallicities from several different parameters, all referred
to the integrated parameter Q$_{39}$, and tied to a high resolution
spectroscopic scale based on old photographic echelle spectra (see Zinn \&
West 1984). Moreover, there have been in the past years several claims (e.g.
Manduca 1983; Frogel et al. 1983) that the integrated light measurements, upon
which the ZW scale was primarily based, are likely to underestimate the true
metallicities of clusters with exceptionally blue horizontal branches (HBs),
since the HB morphology affects the determination of the Q$_{39}$ index.

Very recently, Carretta \& Gratton (1997; CG97) used high dispersion, high
signal-to-noise spectra of more than 160 red giants in 24 clusters to derive a
new metallicity scale based on direct detailed abundance analysis, coupled with
the most recent and upgraded model atmospheres (Kurucz 1992). The
observational material consisted in equivalent widths (EWs) measured on high
quality CCD echelle spectra. Different sets of EWs taken from literature were
carefully checked and, if necessary, brought on a common, homogeneous system,
tied to EWs from the highest resolution spectra. The same set of atomic line
parameter was used for all stars, with model atmospheres from the Kurucz grid.
Also the reference value for the solar [Fe/H] was obtained from the same set
of atomic parameter and the solar model extracted from the same grid used for
giant stars. Input atmospheric parameters (effective temperature and gravity)
for all stars were obtained from the Frogel et al. papers (e.g. Frogel et al.
1983, but see detailed references in CG97), mostly based on accurate infrared
colours and magnitudes. The choice of a temperature scale has an impact on the
derived metallicities: CG97 estimate that the adopted one cannot be
systematically incorrect by more than about 50 K, given the small differences
found in abundances derived from \ion{Fe}{i} and \ion{Fe}{ii}. They also
estimate that random errors, due to uncertainties in individual star colours
and cluster reddenings,  are of the same order of magnitude.

The average internal uncertainty in metal abundance on the CG97 scale is 0.06
dex, resulting in a precise ranking of cluster metallicities. CG97 also
demonstrated that ZW's scale is clearly non-linear, in comparison to their
improved scale. All [Fe/H] values on the ZW's scale were then translated to
the CG97 scale by means of a quadratic interpolating relation, covering the
range in [Fe/H] spanned by the 24 calibrating clusters ($-2.5 \lsim $ [Fe/H]
$\lsim -0.5$).

In the present paper we recalibrate some of the most used metal abundance
indicators to the new CG97 scale, concentrating  on those  based on the
morphology and position of the red giant branch (RGB) in the colour-magnitude
diagram (CMD). We will present calibrations for the indices \bmvg, $\Delta
V_{1.2}$, \d11 in the $V,B-V$ plane, and the analogous in the $V,V-I$ one,
as defined in Section 3.

Such a revised calibration is needed, since the advent of sophisticated high
resolution imaging facilities outside the atmosphere, like  the {\it Hubble
Space Telescope}, allows the observations of clusters in the whole Galaxy,
providing CMDs with giant branches well defined also for very distant or
obscured objects. Precise photometry of extragalactic clusters (in M31,
in the Magellanic Clouds and in Fornax) is also feasible (see e.g. Fusi Pecci
et al. 1996), and it is possible to derive for them quite accurate
metallicities, provided that a good calibration from nearby clusters is
available.

\section{Photometric metallicity indicators in the V,B-V plane}

The position and morphology of the RGB in the $V,B-V$ plane are theoretically
well tied to the metal content of the stars in a cluster: the higher the metal
content, the cooler the effective temperature T$_{eff}$ and the redder the RGB
stars.

In principle, for every GC with a good CMD of the brightest evolutionary
phases, metallicity indicators may be derived from its RGB. However, to obtain
a reliable calibration, homogeneous measurements are needed. We will then
select data sets homogeneous enough to match the quality of the calibrating
metallicity scale.

\subsection{The $(B-V)_{0,g}$ index}

The index $(B-V)_{0,g}$ (Sandage \& Smith 1966) is the de-reddened colour 
of the RGB at the luminosity level of the HB in the $V,B-V$ CMD.

As one of the most homogeneous available samples, we adopted the one published
by Sarajedini and Layden (1997, SL). Their study extended to the $V,B-V$ plane
the Simultaneous Metallicity Reddening method by Sarajedini (1994). They
selected high quality CCD photometric studies of 17 globular clusters (15
galactic and 2 Magellanic Clouds clusters), 6 of which were used as primary
calibrators. De-reddened colours are adopted from their tab. 5.

Whenever possible, we try to rest our calibration on the GCs directly analyzed
in CG97. Among the 17 SL clusters, only 9 have [Fe/H] values from direct
high-resolution spectroscopy; for all other cases, we translated the older ZW
value to the new scale with eq. 7 in CG97. The range in metallicity covered by
CG97 is $-2.5 \lsim $ [Fe/H] $\lsim -0.5$, so all relations here derived are
strictly valid only in this interval. We did not try to extend their validity
to higher metallicities, e.g. by applying a constant offset given by the
difference between ZW and CG97 values at [Fe/H] = $-0.54$ (the most metal-rich
clusters in common): there is in fact the possibility that the strong Ca
lines, upon which ZW determinations are based, saturate at very high
metallicities.
 
SL used the old ZW scale (ZW; Da Costa \& Armandroff 1990; Armandroff et al.
1992). To derive the two expressions for [Fe/H], as a function i) of
$(B-V)_{0,g}$ and ii) of $\Delta V_{1.2}$ needed to simultaneously solve for
metallicity and reddening, they fitted the data using linear regressions. Note
however that in previous studies (like Costar \& Smith 1988) the linearity of 
the $(B-V)_{0,g}$ calibration at the low and high metallicity ends was
questioned. We have repeated the calibration, but in terms of [Fe/H]$_{CG97}$,
and results are shown in Figure 1. Adopting the CG97 scale, this non-linear
effect is obviously enhanced, and it is possible to see also by eye that a
linear fit is a poor approximation.

The resulting best-fit quadratic relations connecting \bmvg \ 
and [Fe/H]$_{CG97}$ shown in Figure 1, upper panel, are:
$${\rm [Fe/H]} = 20.103 (B-V)_{0,g} -9.141 (B-V)^2_{0,g}  \\
                 -11.595 \hskip.1cm(1) $$
\noindent
when using only the 6 SL primary calibrating clusters (with 
$r.m.s.$ deviation $\sigma=0.067$, and correlation coefficient $r=0.997$) and
$${\rm [Fe/H]} = 20.129 (B-V)_{0,g} -9.253 (B-V)^2_{0,g} 
                -11.532 \hskip.1cm(2)$$
\noindent
when using all the 17 SL clusters ($\sigma=0.059$, $r=0.994$). Error bars in
[Fe/H] can be derived from the CG97 paper: they range from 0.01 to 0.11~dex,
with an average value of 0.06~dex. SL apparently did not quote any error
associated to their $(B-V)_{0,g}$ values.

To corroborate the visual impression of non linearity, we tested the
statistical significance of the terms of higher order in Eqs. 1 and 2
by a $t$-test.

The lower panel of Figure 1 displays instead the calibration 
based only upon the 9 CG97 reference clusters; the corresponding 
relation is:
$${\rm [Fe/H]} = 21.790 (B-V)_{0,g} -10.155 (B-V)^2_{0,g} 
		-12.262 \hskip.1cm (3) $$
\noindent
($\sigma=0.041$ and $r=0.998$). In our view, Eq. 3 is the best
interpolating fit, since it has the lower formal statistical $r.m.s.$ and
higher correlation coefficient. Note however that differences in the derived
[Fe/H] values from the one obtained from Eqs. 1 and 2 look negligible (0.01 to
0.02~dex on average, on a range of 0.35 mag in colour).

As a check for the validity of the relations found, we used $(B-V)_{0,g}$
values from two recent high quality photometric studies, namely M3
($(B-V)_{0,g}$ = 0.80; Ferraro et al. 1997) and M5 ($(B-V)_{0,g}$ = 0.83;
Sandquist et al. 1996), which are not among the clusters used to derive the
calibrations. Using Eq. 3 we obtain [Fe/H]$=-1.33$ for M3, and [Fe/H]$=-1.17$
for M5. These values have to be compared with $-1.34 \pm 0.06$ (M3) and $-1.11
\pm 0.11$  (M5) obtained from direct analysis by CG97.

This test suggests that with the present calibration we are able to establish
a very good ranking in cluster metallicities, quite comparable with that given
by the CG97 scale. Eq. 3 comes out as best calibration of the $(B-V)_{0,g}$
parameter as metallicity indicator, and can be adopted as one of the basic
equations of the Simultaneous Metallicity Reddening  method (SMR, Sarajedini
1994).

\begin{figure*}
\hspace{3cm}\resizebox{12cm}{17cm}{\includegraphics{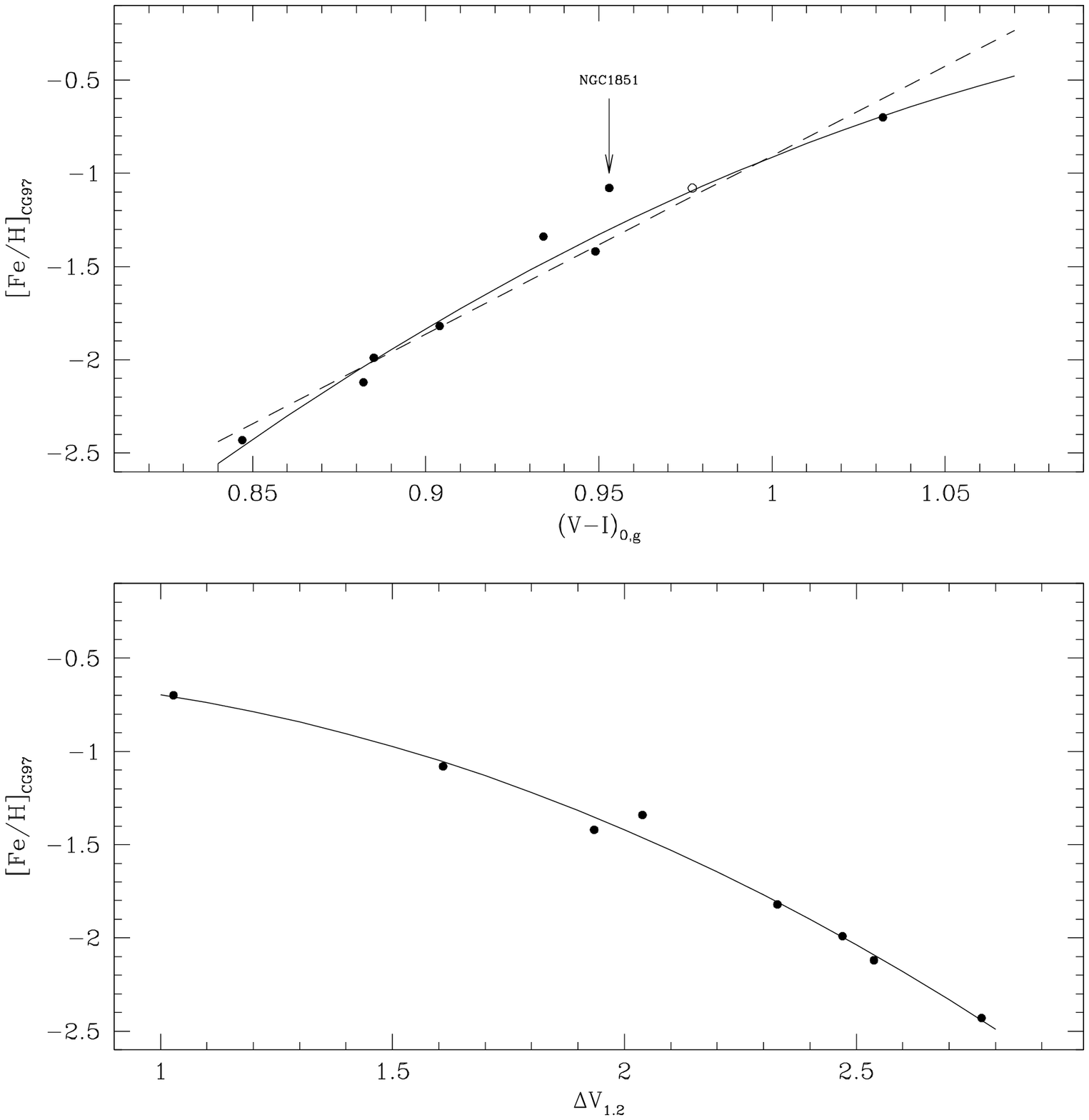}}
\caption{Calibration of the $(V-I)_{0,g}$ (upper panel) and $\Delta V_{1.2}$
(lower panel) parameters in the $V,V-I$ plane using the new CG97 metallicities.
The solid lines in the both panels represent a quadratic interpolation through
the data; the dashed line is the linear best fit. In the upper panel the open
symbol represents NGC1851 once "corrected" as explained in the text.}
\end{figure*}

\subsection{The $\Delta V_{1.2}$ and $\Delta V_{1.1}$ parameters}

The second index we recalibrated is a variation of the classical $\Delta
V_{1.4}$ parameter, that measures the difference in $V$ magnitude between the
HB and the level of the RGB at the de-reddened colour $(B-V)_0=1.4$ (Sandage
\& Wallerstein 1960).
For consistency, we used again the data set from SL, that measured instead the
indices $\Delta V_{1.2}$ and $\Delta V_{1.1}$, referred to the de-reddened
colours $(B-V)_0=1.2$ and $(B-V)_0=1.1$, respectively. As stated by SL,
choosing bluer reference colours could be useful in the case of RGBs poorly
populated in their upper parts.

These parameters, taken as before from  SL tab. 5, have been calibrated (see
Eqs. 4 and 5), and the case of $\Delta V_{1.2}$ is presented in Figure 2
($\Delta V_{1.1}$ has a very similar behaviour and is not shown).

At odds with the case of \bmvg, there seems to be a clear difference between
the calibrations based on the 6 SL primary calibrating clusters and on the
whole SL sample. In particular, for a given $\Delta V_{1.2}$ the first
relation gives a lower value of [Fe/H], and the effect seems to be stronger at
low/intermediate metallicity, while at higher metallicities the two lines
intersect. In fact, using the relation derived from the 6 SL calibrators
[Fe/H] values are underestimated on average by 0.08~dex in the interval
$-1.9 \lsim $ [Fe/H] $\lsim -1.0$, with respect to the other calibration.

We have no explanation for this feature; here we only want to note that SL
stated that {\it ``These secondary calibrators have not been used in the
determination of the (omissis) fitted relations. They only serve to
corroborate these relations''}. It is difficult to see from their fig. 8 if
also with ZW metallicities their primary calibrating clusters provide an
underestimation of [Fe/H]. It is however interesting to note that the
application of the SMR method in the $V$,$B-V$ plane resulted in lower derived
abundances with respect to spectroscopic determinations (based e.g. on the
\ion{Ca}{ii} triplet).

Finally, we re-calibrated the $\Delta V_{1.2}$ - metallicity relation using
the 9 CG97 primary calibrating clusters. The relation is shown  in the lower
panel of Figure 2 and is given by:
$${\rm [Fe/H]}  = 0.236 \Delta V_{1.2} -0.245 \Delta V^2_{1.2}
		      -0.627 \hskip1.7cm(4)$$
\noindent
($\sigma=0.072$, $r=0.993$). 
This calibration provides [Fe/H] values  agreeing very well with those
obtained from all the 17 clusters of SL, and clearly represent a good fit to
all the data.

The case for the $\Delta V_{1.1}$ index closely reproduces that of $\Delta
V_{1.2}$; the resulting calibration based on the 9 CG97 clusters is:
$${\rm [Fe/H]} = 0.089 \Delta V_{1.1} -0.248 \Delta V^2_{1.1}
	        -0.612 \hskip1.7cm(5)$$
\noindent($\sigma=0.075$, $r=0.992$). 

In conclusion, these new calibrations are able to provide metal abundance with
a $r.m.s.$ dispersion of about $0.04 \lsim r.m.s. \lsim 0.07$ dex,
comparable to the errors usually obtained from direct, high resolution
spectroscopy of stars in GCs.

Eqs. 3 and 4 (or 3 and 5) can be used in  the application of the SMR
method in the $V,B-V$ plane.

\section{Re-calibration of the SMR method in the V,V-I plane}

The SMR method was originally devised by Sarajedini (1994) in the $V,V-I$
plane. His calibration was based on the ZW scale and high precision CCD
photometry obtained by Da Costa \& Armandroff (1990).

For the present analysis we used values for 6 clusters as in tab. 1 of
Sarajedini (1994), integrated by values for NGC 5053 (Sarajedini \& Milone
1995) and M 68 (from photometry of Walker 1994, as quoted in Sarajedini \&
Milone 1995) in order to extend the method to very low metallicity clusters.
Data are shown in Table 1.
The parameter definition is analogous to that in the $V,B-V$ plane: we
want [Fe/H] as a function both of $(V-I)_{0,g}$ and of $\Delta V_{1.2}$.

Due to the small sample, all clusters were used to obtain the new
calibrations. Original values from direct analysis by CG97 were used whenever
possible; for the 3 remaining GCs, we transformed ZW values to the CG97 scale.

Figure 3 shows the resulting calibrations (and also a potential problem for
the $(V-I)_{0,g}$ index). The relation obtained using the 8 clusters for
$\Delta V_{1.2}$ is well fitted with a quadratic polynomial (Figure 3, lower
panel), like for the $V,B-V$ plane, and is given by:
$${\rm [Fe/H]} = 0.304 \Delta V_{1.2} -0.342 \Delta V^2_{1.2}
		      -0.659 \hskip1.8cm(6)$$
($\sigma=0.064$, $r=0.996$). 
The quadratic term is found to be highly significant, with a confidence level 
of 99.5 \%.

In the case of $(V-I)_{0,g}$ instead, both a linear and a quadratic
interpolation seem to fit equally well the data (see Figure 3, upper panel).
The statistical significance of the quadratic term is lower than in all
previous cases, being only between 90 and 95 \%. Circumstantial evidence in
favour of a quadratic relation comes, in our view, from the fact that the
relation between $(V-I)_{0,g}$ and $\Delta V_{1.2}$ seems to be a straight
line (see Figure 4), the only discrepant point being NGC1851, and the fact
that $\Delta V_{1.2}$ $is$ quadratically related to [Fe/H].

We have derived again the calibration for the $(V-I)_{0,g}$ index in the
supposition something is wrong with the $(V-I)_{0,g}$ value for NGC1851. We
have  ``corrected'' it to the value it should have if it followed the fit in
Figure 4 (a correction of 0.024 mag, rather large given the errors quoted by
Da Costa \& Armandoff 1990). The statistical significance of the quadratic
term remains virtually unchanged even if, of course, the $r.m.s.$ dispersion of
the quadratic interpolation decreases (from 0.12 to 0.08~dex).

As a further test, we used the value $(V-I)_{0,g}$ = 0.951 (Ferraro et al.
1997) for M3, and obtained [Fe/H]$=-1.27$ and $-1.34$ using the quadratic and
linear interpolations with the original $(V-I)_{0,g}$ value for NGC1851,
respectively; and $-1.32$ and $-1.37$ using the equivalent relations where the
$(V-I)_{0,g}$ has been corrected as described. These values, while nicely
bracketing the spectroscopic value [Fe/H]$=-1.34 \pm 0.02$ (CG97), do not tell
anything definitive about the true form of the calibration at the high
metallicity end.

In view of this, we give both the linear and quadratic expressions for
the $(V-I)_{0,g}$ -- [Fe/H] relation (with the modified $(V-I)_{0,g}$ value for
NGC1851):
$$ {\rm [Fe/H]} = 9.586 (V-I)_{0,g} -10.491 \hskip3.2cm(7) $$
$$ {\rm [Fe/H]} = 42.981 (V-I)_{0,g} -17.772 (V-I)_{0,g}^2
 -26.122 \hskip.1cm(8)$$
We however think it safer to apply the SMR method in the $V,B-V$ plane, at
least until more homogeneous values for $(V-I)_{0,g}$ are supplied.

\begin{figure}
\hspace{.75cm}\resizebox{8cm}{8cm}{\includegraphics{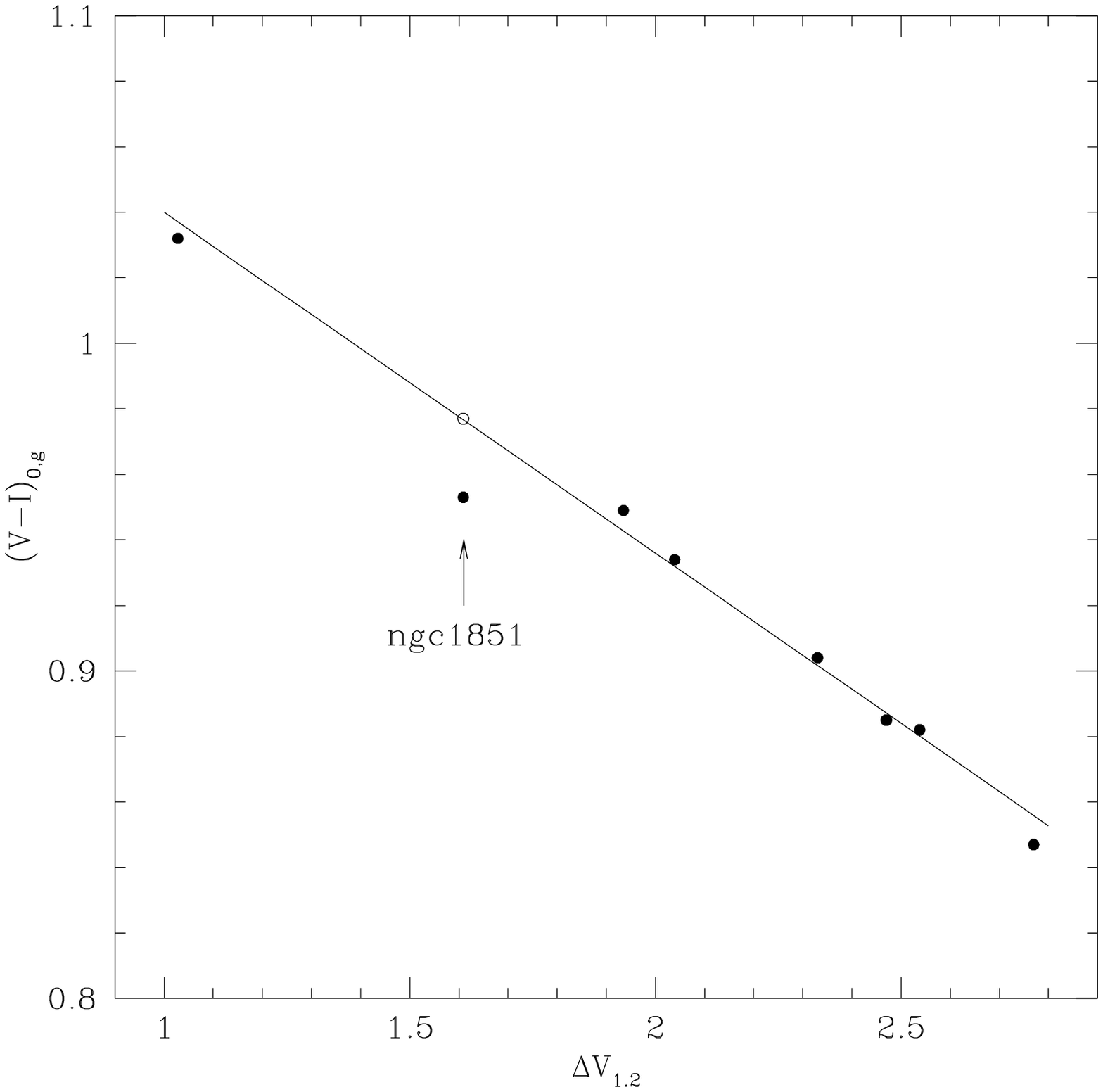}}
\caption{ Comparison of the $(V-I)_{0,g}$ and $\Delta V_{1.2}$ values used
by Sarajedini (1994) and Sarajedini and Milone (1995) to calibrate the SMR
method in the $V,V-I$ plane. The open symbol represents the value of 
$(V-I)_{0,g}$ for NGC1851, "corrected" as explained in the text.}
\end{figure}

Moreover, note that while the $\Delta V_{1.2}$ values in the $V,V-I$ and
$V,B-V$ planes are linearly correlated, the run of $(V-I)_{0,g}$ against
$(B-V)_{0,g}$ is not so clear, especially at the high metallicity end.
Uncertainties in the transformation between the absorption coefficients in
different bands, random errors in the adopted reddenings, the adoption of
different standard system for the $I$ magnitudes could all be possible sources
for the observed disagreement, that seems to affect the $(V-I)_{0,g}$, but much
less a differential measure as the $\Delta V_{1.2}$.

\section{Summary}

We used the new and  homogeneous metallicity scale, derived by CG97 from
updated model atmospheres and direct detailed abundance analysis of high
resolution spectra of globular cluster giants, to calibrate the traditional
RGB photometric indicators in terms of the [Fe/H] ratio.

Especially in the $V,B-V$ plane, the relations found provide very good
relative determinations of [Fe/H] with an uncertainty  on a single measurement
of 0.08~dex on average, and as low as 0.04~dex. This goes towards lessening
the disagreement existing in the past between photometric and spectroscopic
determination of the metal abundance in globular clusters.

\begin{acknowledgements}
EC wishes to dedicate this work to the memory of his beloved grandmother,
Maria Donata.
\end{acknowledgements}

\end{document}